\documentstyle[aps]{revtex}
\begin{document}

\begin{center}
{\bf SINGLE TOP FROM TECHNIPION PRODUCTION}
\end{center}

\begin{center}
PHILIP BARINGER
\end{center}

\begin{center} 
Department of Physics and Astronomy, University of Kansas
\end{center}

\begin{center}
Lawrence, KS 66045
\end{center}

\begin{center}
PANKAJ JAIN
\end{center}

\begin{center} 
Indian Institute of Technology, Kanpur, 208016, UP, India
\end{center}

\begin{center}
DOUGLAS W. MCKAY 
\end{center}

\begin{center}
Department of Physics and Astronomy, University of Kansas, 
\end{center}

\begin{center}
Lawrence, KS 66045 
\end{center}

\begin{center}
LESLEY L. SMITH
\end{center}

\begin{center}
Department of Physics and Astronomy, University of Kansas,
\end{center}

\begin{center}
Lawrence, KS 66045
\end{center}

\begin{center}
{\bf Abstract}
\end{center}
We investigate the contribution of technicolor mechanisms to the production of 
single top quarks at hadron colliders.  
Technipions with mass larger than the top mass will decay predominantly to a top
quark plus a bottom antiquark.  We investigate two promising sub-processes: 
technipion plus W-boson via gluon-gluon fusion and
technipion plus quark production via quark gluon interaction.
The decay 
chain of technipion to top plus bottom quarks and then top to W plus bottom 
yields  final states for the two subprocesses
 with, respectively, 
two W's and two bottom quarks and one W, two bottom quarks and a light
quark.  
We calculate the total cross sections and the $p_{T}$ distributions for these
technipion production mechanisms at Tevatron and LHC energies for a range 
of technipion masses, starting at 200 GeV.
We study the 
backgrounds to our processes and the kinematic cuts that enhance the signal
to background ratio and
we report event rate estimates for the upgraded Tevatron and the LHC.
Only the LHC has the potential to observe these processes.
\newpage
\section{Introduction} 
The discovery of the top quark \cite{CDF/d0} has opened up 
an exciting area of physics.  
Non-Standard Model (SM) production of top \cite{Eichten}, or decay of top 
\cite{Kane}
or a combination of the two \cite{Eichten2}
 have received considerable attention.  
To implement the dynamical symmetry breaking alternative to the SM, research 
continues  in pursuit of complete models that realize the technicolor idea
\cite{Weinberg}
 that there is a new strong force that drives the formation of an 
electroweak symmetry breaking condensate \cite{scenarios}.
  One variation on this theme
explicitly invokes new top physics in conjunction with technicolor to make
 progress on this dynamical symmetry breaking problem \cite{Hill}.
  In partnership
with model building, calculations that produce tests of features that are 
generic to a class of models at active or planned accelerators are needed.  
This paper aims at contributing to such tests in the realm of single top 
production, an area of intense interest for Tevatron and LHC physics 
\cite{Willenbrock}.
\par
In one form or another, an extended technicolor scenario seems necessary for
producing quark masses in a technicolor picture of electroweak symmetry 
breaking.  Extended technicolor generally implies that technipions, like the SM
Higgs, couple to fermions with strengths proportional to the fermion masses.
A positive technipion, for example, should decay to $t {\overline b}$.  In
the SM single top production in a hadron collider requires an extra W-boson 
interaction, and a corresponding factor of $\alpha_{2}$ in comparison to
$t {\overline t}$ production, the production rate of single t 
at the LHC
is expected to
be relatively 
small, $\simeq$ 312 pb \footnote{This value is the sum of single top plus single
anti-top production cross sections.}
 \cite{Heinsohn}.
  These considerations lead us to 
consider production of a single, colored charged technipion and its subsequent
decay into $t {\overline b}$ or $ {\overline t} b$, where the SM top
backgrounds may be manageable and the gluon couplings and extra color factors
help to enhance the technipion production.
\par
Clearly, since production of a neutral technirho with subsequent decay to a pair
of charged technipions would be much larger than our processes, we are {\it not}
proposing a new ``discovery channel". We are looking for a discriminator among 
scenarios where color octet, charged technipions occur, which covers variants
of the one-family models that accomodate precision electroweak constraints
\cite{Goldman}, including ``top-color-aided" cases 
\cite{Hill}.  As we will see below, 
the single technipion production involves anomalous terms in the chiral
effective Lagrangian and distinctive kinematics of the final state decay 
products.  These features make background suppression possible, and test 
interesting features of the technicolor models.
\par
In the next section we describe the calculation and present representative 
cross-section plots.  In Section 3, we discuss backgrounds and ways in which a
signal may be observed.  In Section 4 we discuss our results 
and 
conclude.  An appendix gives detailed formulas for cross-sections referred to
 in the text.
\section{Single Top Produced from Technipion Decay}
Figure 1 shows the subprocesses that yield a single, charged,
color-octet technipion, $P_{8}^{+}$.  The $t \overline{b}$ mode is assumed
 to dominate the $P_{8}^{+}$ decay, so these are our technicolor-driven,
single top production mechanisms.  The direct $P_{8}^{+}$ production from 
the heavy quark sea, Figure 1c, is an order of magnitude smaller than the
gluon-gluon fusion process, Figure 1a, and two orders of magnitude smaller
than the gluon-quark
process, Figure 1b, so we do not discuss it further in this paper.
\par
In Figure 2, we show the Feynman graphs that contribute to $P_{8}^{+}$
production in lowest order in $g_{3}$ and $g_{2}$, the QCD and weak SU(2) 
couplings.  The technipion exchange term in Figure 2b makes an insignificant
contribution because the technipion Yukawa couplings are taken to be
proportional to the fermion masses they couple to, which is 
generally the case in extended technicolor scenarios.
The terms shown in the diagrams in Figure 2b are individually gauge invariant.
\par
In Figure 2 the heavy dots indicate anomalous vertices.  The relationship
between the four-point $ggW P_{8}^{+}$ anomalous vertex and the 
three-point $gW P_{8}^{+}$ vertex in Figure 2a is fixed by gluon gauge
invariance and is therefore model independent.  The overall factor for the 
anomalous vertex, and therefore the overall factor for the amplitudes in Figures
2a and 2b {\it is} model dependent.  As is evident from the diagrams, 
the factor has the form $\kappa_{a} g_{3}^{2} g_{2}$ in Figure 2a and
$\kappa_{b} g_{3} g_{2}^{2}$ in Figure 2b.  In the one-family technicolor
model, for example, $\kappa_{a}=\frac{i N_{T}}{16 \pi^{2} F_{T}}=\kappa_{b}$,
where $SU(N_{T})$ is the 
technicolor group and $F_{T}=125$ GeV is determined by the
weak scale \cite{Farhi}.
\par
The processes shown in Figure 2 have distinct signatures.  The
$P_{8}^{+} W^{-}$ 
will have a high $p_{T}$ recoil $W^{-}$ to help tag the events.  
Similarly, the quark jet recoiling against $P_{8}^{+}$ in the 
$P_{8}^{+} q$ 
final states will provide a high $p_{T}$ jet tag.  
\par
In Figure 3 we show the total cross-sections 
for $pp$ collisions at $\sqrt{s}$=14 TeV for $gg$ and $gq$ (
$P_{8}^{+} W^{-}+P_{8}^{-} W^{+}$  and $P_{8}^{+}d+P_{8}^{-}u$ final states,
respectively)
as a function of the technipion mass, 
$M_{P_{8}^{+}}$, for the one-family technicolor model with
$N_{T}=3$.  We use this model for reference because it is familiar, has the
charged, color-octet technipion necessary for our mechanism, and variants
that have been proposed to avoid conflict with precision electroweak data
also have charged color-octet technipions; only the factors ``$\kappa$"
multiplying the 
anomalous 
vertices are different.  
Therefore, results for a class of alternative models can be 
obtained by multiplying our ``reference model" results by the 
appropriate factor.
\par
The $p_{T}$ distribution of the $P_{8}^{+}$ 
at $y=0$ for each process is shown in Figure 4.   We
 take $M_{P_{8}}=3M_{W}$.  In the $P_{8}^{+} W^{-}$
case, the $p_{T}$ is equivalent to 
that of the $W^{-}$, while in the $P_{8}^{+} d$
 case, the $p_{T}$ is equivalent to 
that of the quark jet.  The $p_{T}$ spectra are quite 
stiff because of the extra momentum factors appearing at the anomalous coupling,
as was noted in earlier work on similiar technicolor processes \cite{MYS}.
\par
The numbers of events for ${\cal L}=100 fb^{-1}$ 
corresponding to LHC parameters and $10^{7}$ seconds, or 
one-third of a year (a ``Snowmass year")
 of running, are shown for the two  $P_{8}^{+}$ production
processes in Tables 1 and 2 
for a technipion mass of 240 GeV.
The two charge states are added together, and the reference
model has been used ({\it i.e.}, one family technicolor with $N_{T}=3$).
We will comment in Section 4 on the numbers obtained in other, perhaps more 
realistic, models.

\begin{table}
\caption{ The total cross section, $\sigma$, the number of events in one 
``Snowmass year" ($10^{7}$s) of running at
LHC design luminosity times
$\sigma$, $N$(total), and the number of events that remain after
cuts, $N$(after cuts), for the $WWb \overline{b}$ final state. (See Sec. 3
for cuts.) }

\begin{center}
\begin{tabular}{|l|c|c|c|} \hline
	& $\sigma$ & $N$(total) & $N$(after cuts)
\\ \hline
Signal & 0.112 pb & 1.12 $\times 10^{4}$ & 185 \\ \hline
Background & 526 pb & 5.26 $\times 10^{7}$ & 910 \\ \hline
\end{tabular}
\end{center}
\end{table}
\begin{table}
\caption{ As in Table 1 for the $Wb \overline{b} q_{jet}$ final state from
the signal processes 
$pp \rightarrow P_{8}^{+} d X + pp \rightarrow P_{8}^{-} u X$ and the
background process 
$pp \rightarrow t \overline{b} q X + pp \rightarrow \overline{t} b q X$.  
(See Sec. 3 for cuts.)}
\begin{center}
\begin{tabular}{|l|c|c|c|} \hline
  & $\sigma$ & $N$(total) & $N$(after cuts)
\\ \hline
Signal & 0.94 pb & 9.4 $\times 10^{4}$ & 580 \\ \hline
Background & 225 pb & 2.25 $\times 10^{7}$ & 990  \\ \hline
\end{tabular}
\end{center}
\end{table}
\par
Under our assumption that the dominant $P_{8}^{+}$ decay mode is 
$t \overline{b}$, the final states will look like:
\par{\noindent}
(a)
\[ \left.
\begin{array}{lcccr}
P_{8}^{+} & \rightarrow & t & \rightarrow & W^{+} b \\
	&	& \overline{b} &  & \\
W^{-} & & & & 
\end{array} \right\}
W^{+} b \overline{b} W^{-} \]
\par{\noindent}
in the gluon-gluon fusion case, Figure 1a.
\par{\noindent}
(b)
\[ \left.
\begin{array}{lcccr}
P_{8}^{+} & \rightarrow & t & \rightarrow & W^{+} b \\
  & &  \overline{b} & & \\
q & & & &
\end{array} \right\}
W^{+} b \overline{b} jet \]
\par{\noindent}
 in the gluon-quark case, Figure 1b.
The total cross sections for cases (a) and (b) are 0.03 fb and 0.4 fb 
for $p \overline{p}$ collisions at 
$\sqrt{s}=2$ TeV, corresponding to an uncut total of 1 and 14 events in a 
year at the upgraded Tevatron -- a dismal prospect.
\par
  In the next section, we address the crucial background issue
for these two signals at LHC energy and luminosity, where the prospects are
reasonable.
\section{Background from Standard Model Top Production}
The SM total cross-section for $t \overline{t}$ and single top
production are approximately 525 pb and 225 pb \footnote{This value is for
the part of the total single top plus antitop cross-section which leads
to the $Wb \overline{b}q$ final state signature that is a background for our
$q+ g \rightarrow P_{8} + q$ process \cite{onetop}.  The total cross-section
is 312 pb \cite{Heinsohn}.}
 respectively.  These
processes are the dominant backgrounds to our technipion, single top signals.
There are kinematic
 features to the technipion processes that 
suggest cuts to beat down these backgrounds to a level
where a signal, if present in the data, could be pulled out.
\par
We first consider the process of Figure 1a which has a 
$W^{+}W^{-}b \overline{b}$ final state, where one $Wb$ pair comes from top 
decay, the other b from the technipion decay, and the other $W$ recoils against
the $P_{8}^{\pm}$.  The following cuts effectively reduce the $t \overline{t}$
background to a manageable level:
\[
\begin{array}{lccc}
i. & p_{T}~ W_{1} & > & 400~ GeV  \\
ii. & p_{T} ~W_{2} & > & 100 ~GeV \\
iii. & p_{T} ~b & > & 50~ GeV~ ({\rm both}~ b'{\rm s})  \\
iv. & | \eta |_{b} & < & 2~ ({\rm both}~ b'{\rm s})  \\
v. & | \eta |_{W} & < & 2~ ({\rm both}~ W'{\rm s})  \\
vi. & cos \theta_{b \overline{b}} & > & 0 
\end{array}
\]
\par
These cuts are motivated by the following considerations:  The $W$ recoiling
against the $P_{8}^{\pm}$ will have very large $p_{T}$.  The decay products 
of the $P_{8}^{\pm}$ will also have substantial $p_{T}$ and the opening angles
between the decay products will be small in the lab.  The $| \eta |$ cuts
restrict our events to the acceptance region of the LHC detectors.  We assume
that these events will be most readily detected in the ``lepton plus jets"
mode where one of  the two $W$'s decays into a pair of jets and the other into
either $e \nu$ or $\mu \nu$.  The charged lepton and large missing energy will
provide the event trigger and with only one neutrino the top mass can be 
reconstructed.  We assume the $b$ jets have been identified as has the lepton
from $W$ decay.  We do not take into account here the inefficiencies of the 
particle  ID's, but this should affect signal and background in a very similar
manner.
\par
With these cuts, we calculate that in one LHC year (100 $fb^{-1}$) there will 
be 185 signal events and 3180 $t \overline{t}$ background events
\cite{onetop}.  Additional discrimination can be achieved by requiring that one 
$Wb$ pair {\it not} reconstruct to the top mass.  The effectiveness of this 
cut will depend upon the mass resolution of the detector.  Using a conservative
resolution assumption, we insist that one $Wb$ mass be at least 200 GeV
and find that the number of background
$t \overline{t}$ events is reduced to 910 events
with negligible effect on the signal.
See Table 1 for signal processes
 $pp \rightarrow P_{8}^{+} W^{-} X + pp \rightarrow P_{8}^{-} W^{+} X$ and for
the background process $pp \rightarrow t \overline{t} X$.  
In addition to the cuts $i.$--$vi.$ listed above, the invariant
mass of one $Wb$ combination was restricted to $|M_{Wb}|>200$ GeV, as just 
mentioned.  
We also considered the non-resonant $WWbb$ background and found it to be 
negligible in comparison with the $t \overline{t}$ background.
\par
In considering the second process from Figure 1b, the chief difference is that
 we have a light quark jet, $q$, rather than a $W$ recoiling against the 
technipion. The dominant background is single top production from $W$-gluon
fusion, giving $Wbbq$ final states with one $Wb$ from top decay.  With just
one $W$ we must have it decay to $e \nu$ or $\mu \nu$ to facilitate the event
trigger.  We replace cut $i.$ above with a cut of 500 GeV on the most energetic
jet, and use the other cuts as before.  This yields 580 signal events,
990 background events from single top, as summarized in Table 2.  
We note that requiring one 
$Wb$ combination to reconstruct to the top mass will provide additional 
discrimination against non-resonant $Wbbq$.  This is already strongly suppressed
by cut $i$.
\section{Discussion of Results and Conclusions}
We have calculated the cross-sections for the elementary single, 
color-
octet technipion processes shown in Fig. 2, which are generic to techicolor 
models with QCD triplet techniquarks.  The technipion decay to top and
bottom yields a single top, whose decay then leads to a $WWbb$ final state
in process (2a) and a $Wbbq$ final state in process (2b).  The standard model
final states that will be the dominant backgrounds after cuts to our single
technipion signal were analyzed and compared to the signal process for the
``reference" one-family technicolor model with $N_{T}=3$
 \cite{Farhi}.  Tables 1 and 2 show the expected signal and background at the 
LHC for a nominal $10^{7}$ second year of running with $M_{P_{8}}=240$ GeV.
Cuts on the $p_{T}$ and $\eta$ ranges of the final particles, as well as the 
$b \overline{b}$ opening angle, are described in the text.
  A variety of cuts were tried,
 but no attempt at an optimization was made.  
For example, the background falls much faster than the signal as $p_{T}~b$
is raised (cut $iii.$).  We have not played the game of raising the cut to find
if there is a signal to background ratio of 1:1.
Our robust result is that a
signal-to-background ratio 
of roughly 1:2 and a signal sample of order $10^{3}$
events can be achieved.  (The uncut number of events falls by roughly a factor 
of $\frac{2}{3}$ for each added 50 GeV of $M_{P_{8}}$ mass.) 
\par
Our result holds up under variations to the Farhi-Susskind model that we used
for the above estimates.  For example, a recently proposed model that
survives precision electroweak constraints \cite{Goldman} is that of Kitazawa
and Yanagida \cite{Kitazawa}.  The techniquarks are an $SU_{T}(3)$ triplet
in their model, and they comprise the quark sector of a family in the
Standard Model.  The parameter $\kappa$ (see Eq. A.1) is then
$\kappa = \frac{i ~N_{T}}{16 \pi^{2}~F_{Q}}$, where $N_{T}=3$ just as in 
our choice for 
 reference model, but $F_{T}=\sqrt{F_{Q}^{2}+F_{L}^{2}}=125$ GeV.
$F_{Q}$ is the decay constant of the color octet technipions,
which couple to techniquarks but not to technileptons, appropriate for
 our 
problem and $tan \phi = \frac{F_{Q}}{F_{L}}$ is a free parameter in their
model.  Clearly the value of $\kappa$, and thus the production rate of the
color-octet technipions, is enhanced in this model compared to that of the 
reference model.  For a given technipion mass, our estimates are therefore 
on the
conservative side.
\par
An example of a topcolor assisted technicolor model in which color-octet
technipions occur is that of Lane \cite{Lane}. 
 The pseudogoldstone boson content of this
model is rather complex, however, and the diagonalization
 problem to get mass eigenstates and identify those that carry QCD color 
quantum numbers has not been worked out.  As with the model of \cite{Kitazawa},
however, the effective $F$ that enters in the coefficient $\kappa$
 (Eq. A.1) will
be less than $F_{T}=125$ GeV, thus giving an enhancement.
\par
In conclusion we believe that the study presented in this paper shows that
if some variation of technicolor is indicated by dominant processes such as 
technirho production and decay into a pair of technipions, our single 
technipion processes would be observable at the LHC.  The single top
final states would then become an exciting hunting ground for signals
to elucidate the new physics.
\section*{Acknowledgements}
This research was supported in part
by DOE grant \# DE - FG02 - 85ER40214 and NSF grant PHY-9504357.
  The computational facilities of the
Kansas Institute for Theoretical and Computational Science were used 
for part of this work.  

\appendix
\section*{Appendix}
Here we collect the principal components of the calculation reported in the
text.  The relevant, anomaly-driven vertices are given by the effective
Lagrangian \cite{mck}

\begin{eqnarray}
{\cal L}^{A}_{eff} 
& = & i \kappa g_{2} g_{3} \epsilon^{\mu \nu \lambda \rho}
\left( \partial_{\mu} G^{b}_{\nu} + \frac{1}{2} g_{3} f^{bcd}
G^{c}_{\mu} G^{d}_{\nu} \right) \nonumber \\
 & \times & \left( P^{+,b}_{8} \partial_{\lambda} W^{-}_{\rho}
+ P^{-,b}_{8} \partial_{\lambda} W^{+}_{\rho} \right), 
\hspace*{3.75cm}(A1) \nonumber
\end{eqnarray}
where $P^{\pm,b}_{8}$ are the color-octet technipions and $G^{b}_{\mu}$
are gluons.  The constant $\kappa$ has dimension $(M{\rm ass})^{-1}$ and it depends
on the specific model employed  The trilinear gluon-$P^{+}_{8}-P^{-}_{8}$
vertex that follows from the $P_{8}$ kinetic energy is
\begin{eqnarray}
{\cal L}_{G} &=&  -g_{3} f^{abc} \partial^{\mu} P^{+,a}_{8}
P^{-,b}_{8} G^{c}_{\mu}. \hspace*{4.5cm}(A2) \nonumber
\end{eqnarray}

The amplitude for $g^{a}+g^{b} \rightarrow P^{\pm,c}_{8} +W^{-}$, designated
as ``a", is given by the expression
\begin{eqnarray}
T_{a} & = & -\kappa g_{2} g^{2}_{3} f^{cab} 
\epsilon^{\mu \nu \lambda \rho} \left\{ e_{\mu} (q_{a} ) e_{\nu} (q_{b}) \right.
k_{\lambda} e_{\rho} (k) \nonumber \\
 & + & \frac{(q_{a} +q_{b})_{\mu}}{2 q_{a} \cdot q_{b}} \left[ (2 q_{a} \right.
+ q_{b} ) \cdot e(q_{b}) e_{\nu}(q_{a}) - (q_{a} - q_{b})_{\nu} e(q_{a})
\cdot e(q_{b})  \nonumber \\
& - & 
\left.
(2 q_{b} + q_{a} ) \cdot e(q_{a} ) e_{\nu} (q_{b}) \right] k_{\lambda}e_{\rho}
(k) \nonumber \\
& + & \left[
\frac{(q_{b})_{\mu} e^{b}_{\nu} (q_{b}) k_{\lambda} e_{\rho}(k)
(k-p-q_{b}) \cdot  e(q_{a})}{(q_{b} -k)^{2} -m^{2}} \right. \nonumber \\
 & + & \left.
\left.
b \longleftrightarrow a \right] \right\}. \hspace*{8cm}(A3) \nonumber
\end{eqnarray}

The variable labels are: $q_{a}=$ gluon a momentum, $q_{b}=$ gluon b momentum,
$k=W$-boson momentum, $P=$ technipion momentum, and $m$ is the technipion mass.
The $e_{\lambda}$'s are the vector field polarization vectors and $\kappa$ is 
the model-dependent
 constant defined in Eq. (A1).

The parton-level cross-section that follows from the above amplitude can be 
expressed as
\begin{eqnarray}
\frac{d \sigma}{dt} & = & \frac{3}{8} \pi^{2} 
\kappa^{2} \alpha^{2}_{3} \alpha_{2} 
\frac{1}{s^{2}} \left\{ 2 M^{2}_{W} + 4 \left[ ck ~ tw + dk ~ uw 
\right. \right. \nonumber \\
& + & \left( 2 + \frac{tp}{up} + \frac{up}{tp} \right) 
\frac{tw ~ uw}{2s} + \left( - \frac{m^{2}s}{(tp)^{2}} + \frac{up}{tp} 
\right) \frac{(tw)^{2}}{2s} \nonumber \\
& + & \left( -\frac{m^{2}s}{(up)^{2}} + \frac{tp}{up} \right)
\frac{(uw)^{2}}{2s} - \frac{s M^{2}_{W} m^{2}}{up ~ tp}
+ s ~ ck ~ dk \nonumber \\
& + & 
\left. \left.
\frac{tp}{up} \frac{ck ~ tw}{2} + \frac{up}{tp} ~ 
\frac{dk ~ uw}{2} + \frac{tw ~ uw ~ cd}{2} \right] \right\}
\hspace*{2cm}(A4) \nonumber
\end{eqnarray}

The definitions of the factors in terms of $s,~t,~M^{2}_{W}$ and $m^{2}$ are:
\begin{eqnarray}
u & = & M^{2}_{W} + m^{2} - s - t,  ~tw  =  M^{2}_{W} - t,  ~uw  = 
s+ t - m^{2}, \nonumber \\
up & = & s + t - M^{2}_{W},  ~tp  =  m^{2}-t, ~cd  =  - \frac{1}{2s}
+ \frac{m^{2}}{(up ~ tp)}, \nonumber 
\end{eqnarray}
\begin{eqnarray}
ck & = & \frac{tw}{2s} - \frac{s-M^{2}_{W} - m^{2}}{2tp}, \nonumber \\
 dk & = & \frac{uw}{2s} -\frac{(s-M^{2}_{W} - m^{2})}{2up} .
\hspace*{4.25cm}(A5) \nonumber
\end{eqnarray}

For process ``b", namely $ g^{a}+q^{b} \rightarrow P^{c}_{8} + q^{b}$,
we have the amplitude

\begin{eqnarray}
T_{b} &=& i \kappa \frac{g^{2}_{2}g_{3}}{2 \sqrt{2}} \delta^{a,b}
\delta^{c,d} 
\epsilon^{\mu \nu \lambda \rho}
\frac{q_{\mu} e_{\nu}(q) (k-k')_{\lambda} \overline{u} (k') \gamma_{\rho}
(1-\gamma_{5}) u(k)}{(p-q)^{2}-M^{2}_{W}} \hspace*{1cm}(A6) \nonumber
\end{eqnarray}
where the momenta of gluons, incoming quark, outgoing quark and technipion are
$q, k, k'$ and $p$, respectively.  The technipion exchange amplitude in 
Fig. 2b is negligible.
\par
The corresponding parton level cross section is
\begin{eqnarray}
\frac{d \sigma}{dt} = \frac{\pi^{2}}{4} \kappa^{2} \alpha^{2}_{2} \alpha_{3}
\frac{(-t)}{s^{2}} \frac{(s^{2} + u^{2})}{(t-M^{2}_{W})^{2}} ,
\hspace*{4cm}(A7) \nonumber
\end{eqnarray}
where
\begin{eqnarray}
t &= &(k-k')^{2}  =  -2 k \cdot k' \nonumber \\
u & = & (q-k')^{2}  =  -2q \cdot k' \nonumber \\
s & = & (q + k )^{2}  =  2 q \cdot k,
\hspace*{5cm}(A8) \nonumber 
\end{eqnarray}
and quarks are light and treated as massless.
\par
To compute the $p-p$ production cross sections we used the version 3
CTEQ structure
 functions \cite{cteq}.
  Results are shown for set 2, leading order, versions of the
 structure
functions, though we also used set 1, the $\overline{MS}$ to two loop version
as a check.  The differences between the cross-sections and event rates
were at the few percent level, which is inconsequential for our
purposes.  We show the set 2 results in the spirit of staying within a 
purely leading
order calculation.

\newpage
\begin{center}
Figure Captions
\end{center}

\par
\noindent
FIG.1. (a) The $g+g\rightarrow P_{8}+W$ process; (b) The 
$g+q\rightarrow P_{8}+q$ process; (c) The direct sea contribution from top,
$t+\overline{b} \rightarrow P_{8}$.  Curly lines are gluons, wavy lines are
$W$-bosons, dashed lines are technipions and straight, solid lines are quarks.
\vspace*{.75cm}
\par
\noindent
FIG.2. (a) The four Feynman diagrams that contribute to the gluon-gluon fusion
process, $g+g \rightarrow P_{8}+W$; (b) Feynman diagrams for the quark-gluon 
fusion process $g+q \rightarrow P_{8}+q$.  The heavy dots indicate anomalous
vertices -- otherwise the same conventions as Fig.1.
\vspace*{.75cm}
\par
\noindent
FIG.3.  The total cross sections for (a) 
$pp \rightarrow qq \rightarrow P_{8}+W$ and for (b)
$pp \rightarrow qg \rightarrow P_{8}+q$ vs. the technipion mass, $M_{P_{8}}$.
The c.m. energy is 14 TeV and $N_{TC}=3$ in the one-family Farhi-Susskind
model.
\vspace*{.75cm}
\par
\noindent
FIG.4. The differential cross-section $d \sigma /dydp_{\perp}~|_{y=0}$ for
processes (a) and (b) with $M_{P_{8}}=240$ GeV.

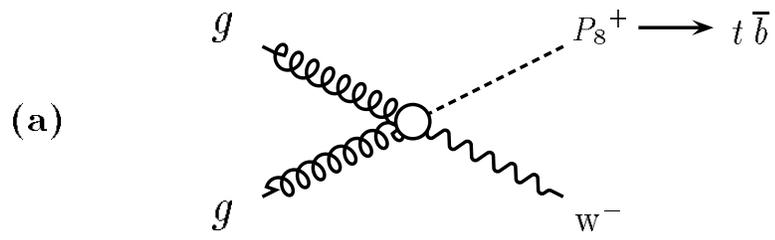
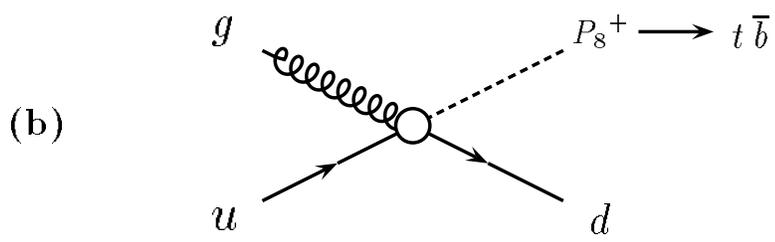
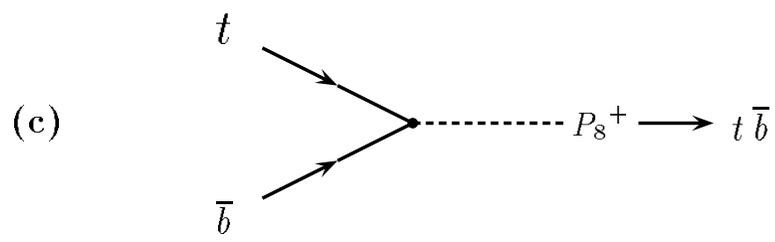
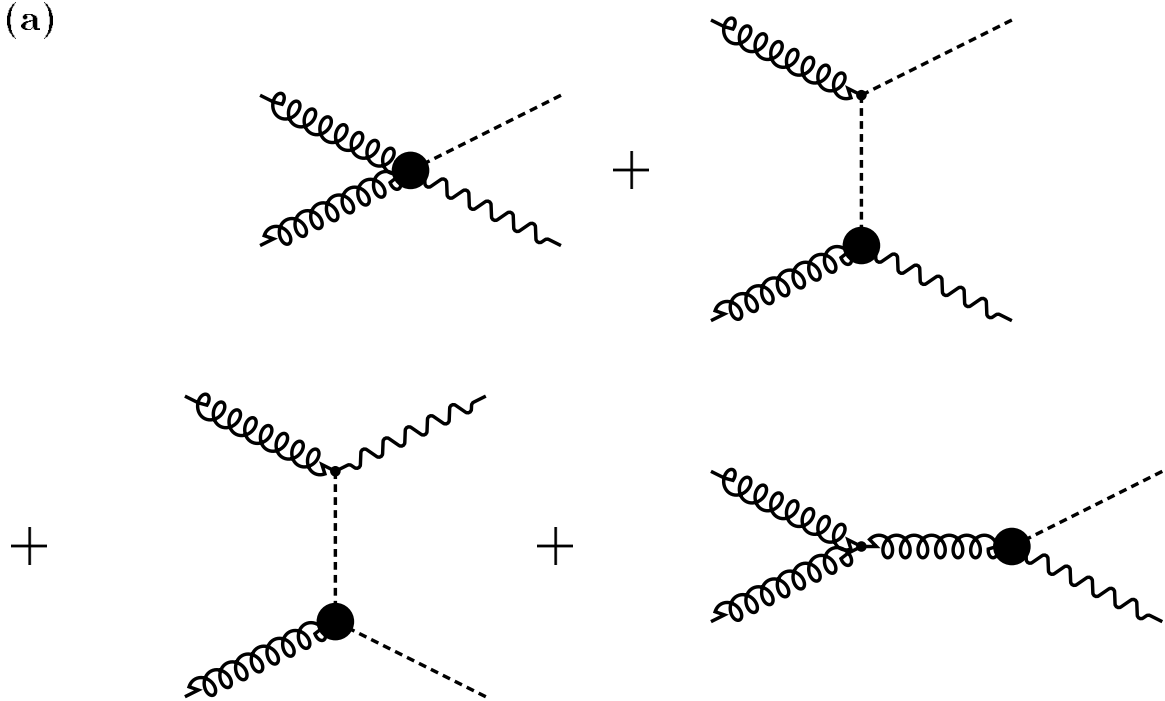
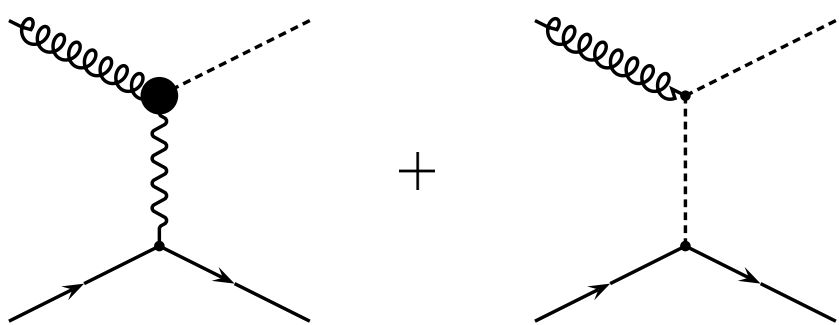
\end{document}